\documentclass[prb,twocolumn,showpacs]{revtex4}

\usepackage{graphicx}
\usepackage{dcolumn}
\usepackage{amsmath}

\newcommand{\GdI}{GdI$_2$ }
\newcommand{\etal}{\textit{et al.~}}
\newcommand{\hres}{$H_{\rm res}$ }
\newcommand{\iesr}{$I_{\rm ESR}$ }

\begin{document}
\title{Spin fluctuations in the quasi-two dimensional Heisenberg ferromagnet GdI$_2$ studied by Electron Spin Resonance}
\author{J.~Deisenhofer}
\affiliation{Experimentalphysik V, Center for Electronic
Correlations and Magnetism, Institute for Physics, Augsburg
University, D-86135 Augsburg, Germany}
\author{H.-A.~Krug von Nidda}
\affiliation{Experimentalphysik V, Center for Electronic
Correlations and Magnetism, Institute for Physics, Augsburg
University, D-86135 Augsburg, Germany}
\author{A.~Loidl}
\affiliation{Experimentalphysik V, Center for Electronic
Correlations and Magnetism, Institute for Physics, Augsburg
University, D-86135 Augsburg, Germany}

\author{K.~Ahn\footnote{Present address: Department of Chemistry,
Yonsei University, Wonju, Korea}} \affiliation{Max-Planck-Institut
f{\"u}r Festk{\"o}rperforschung, D-70569 Stuttgart, Germany}
\author{R.~K.~Kremer}
\affiliation{Max-Planck-Institut f{\"u}r Festk{\"o}rperforschung, D-70569
Stuttgart, Germany}
\author{A.~Simon}
\affiliation{Max-Planck-Institut f{\"u}r Festk{\"o}rperforschung, D-70569
Stuttgart, Germany}

\date{\today}

\begin{abstract}
The spin dynamics of GdI$_2$ have been investigated by ESR
spectroscopy. The temperature dependences of the resonance field
and ESR intensity are well described by the model for the spin
susceptibility proposed by Eremin \etal [Phys.~Rev.~B 64, 064425
(2001)]. The temperature dependence of the resonance linewidth
shows a maximum similar to the electrical resistance and is
discussed in terms of scattering processes between conduction
electrons and localized spins.
\end{abstract}


\pacs{76.30.-v, 71.70.Ej, 75.30.Et, 75.30.Vn}

\maketitle

\section{Introduction}

\begin{figure}[b]
\centering
\resizebox{8cm}{!}{\includegraphics*{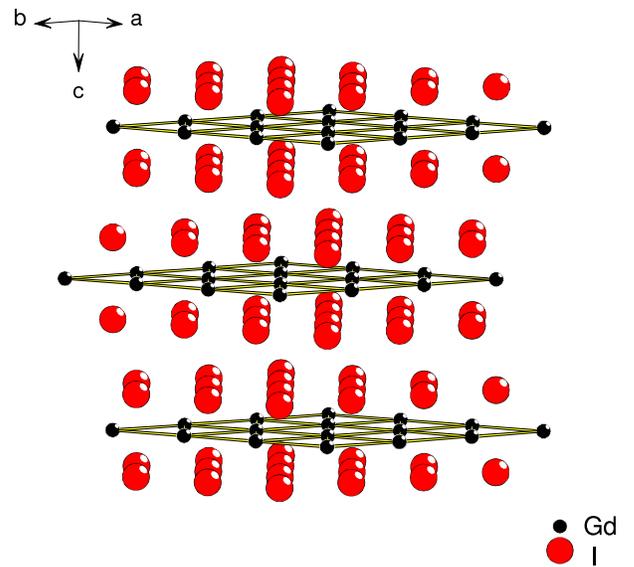}}
\vspace{2mm} \caption[]{\label{struktur}Crystal structure of
GdI$_2$ (space group $P6_3$/mmc).}
\end{figure}

Recently, the phenomenon of giant magnetoresistance has attracted
considerable interest, not only in the manganites\cite{Dagotto01}
but also in spinel-type\cite{Ramirez97} or intermetallic
compounds.\cite{Mallik98a,Baranov01} Gadolinium diiodide came into
the focus of solid-state research because of reports of giant
negative magnetoresistance effects at the ferromagnetic
phase-transition close to room temperature.\cite{Ahn00,Felser99}
GdI$_2$, which contains formally divalent Gd with an electronic
configuration 4$f^7$5$d^1$, is a correlated narrow $d$-band metal
revealing ferromagnetism. \GdI has first been synthesized by Mee
and Corbett.\cite{Mee65} It crystallizes in the close-packed
hexagonal 2H-MoS$_2$ structure (space group $P6_3$/mmc), which has
a strongly two-dimensional (2D) character: the rare-earth layers,
in which each Gd atom is surrounded by six nearest Gd neighbours,
are separated by two I atom layers. The stacking sequence is built
up along the c-axis (see Fig.~\ref{struktur}). A first study of
the magnetic properties was performed by Kasten \etal
characterizing \GdI as a ferromagnet with a Curie temperature near
room temperature,\cite{Kasten84} at which the Gd moments are
aligned perpendicular to the c-axis.

Using a semi-phenomenological approach, the large
magnetoresistance has been explained in terms of a $d$-$f$
exchange model: The anomalous peak of the resistance and the
negative magnetoresistance are due to a strong scattering of the
$d$-derived conduction electrons by the localized 4$f$
electrons.\cite{Eremin01} Furthermore, also from its magnetic
properties \GdI is highly interesting. Gd exhibits a half-filled
$4f$-shell with a spin-only moment of 7 $\mu_{\rm B}$. Any
spin-orbit coupling is expected to be weak and in the paramagnetic
phase \GdI should be a good realization of a 2D-Heisenberg
ferromagnet. Theoretically, a 2D Heisenberg ferromagnet is
expected to order at 0 K only. Of course, interplane interactions
cannot be neglected and \GdI reveals bulk ferromagnetic order
below 290 K with an ordered moment of 7.33 $\mu_{\rm B}$. The
excess moment of 0.33 $\mu_{\rm B}$, when compared to the
spin-only moment of the Gd 4$f$ shell, obviously results from a
strong polarization of the 5$d$ conduction band.\cite{Ahn00}

\section{Sample preparation and experimental details}

\GdI has been prepared in a solid-state reaction of GdI$_3$ and Gd
metal powder at 1100 K for three weeks in a sealed Ta tube
jacketed with an evacuated silica ampoule. Details of the sample
preparation are given in references \onlinecite{Felser99} and
\onlinecite{Ahn00}. The structure has been determined by x-ray
powder diffraction and lattice parameters $a$ = 0.40775(4) nm and
$c$ = 1.5041(1) nm were obtained. For the samples under
investigation, from an analysis using modified Arrotts plots the
ferromagnetic phase-transition temperature has been determined as
$T_{\rm C}$ = 276 K.\cite{Felser99}

ESR measurements were performed in a Bruker ELEXSYS E500
CW-spectrometer at X-band frequencies ($\nu \approx$ 9.47 GHz)
equipped with a continuous N$_2$-gas-flow cryostat in the
temperature region $80<T< 600$ K. Note that the samples are very
sensitive to exposure to air and humidity. Therefore the whole
procedure of powdering the polycrystalline samples and placing
them into quartz tubes had to be undertaken in He atmosphere. Then
the tubes were sealed in He atmosphere, too. ESR detects the power
$P$ absorbed by the sample from the transverse magnetic microwave
field as a function of the static magnetic field $H$. The
signal-to-noise ratio of the spectra is improved by recording the
derivative $dP/dH$ using lock-in technique with field modulation.

\section{Experimental Results and Discussion}
\subsection{ESR spectra}
\begin{figure}[ht]
\centering
\includegraphics*[width=70mm,clip,angle=0]{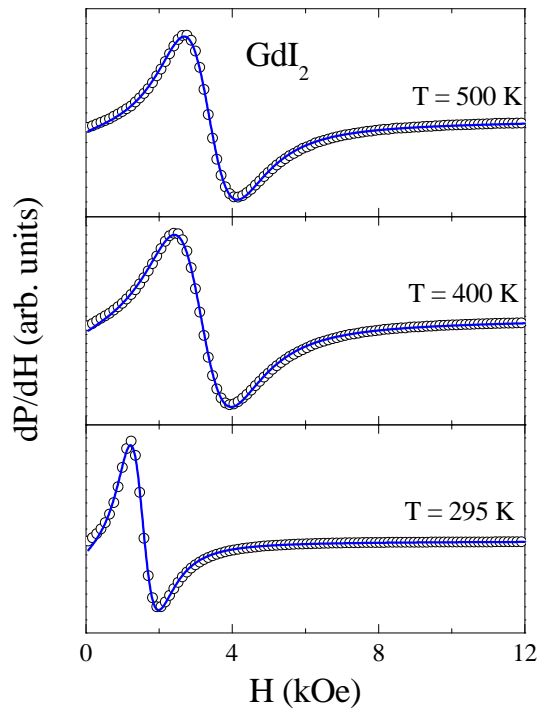}
\vspace{2mm} \caption[]{\label{Spectra}ESR spectra of \GdI for
different temperatures illustrating the line shift and linewidth
changes. Solid lines are fits obtained by using a Dysonian
lineshape according to Eq.~(\ref{dyson}).}
\end{figure}

ESR spectra obtained for GdI$_2$ in the paramagnetic regime at
different temperatures  are displayed in  Fig.~\ref{Spectra}. The
spectra consist of a broad, exchange-narrowed resonance line,
which is well fitted by a single Dysonian line
shape.\cite{Barnes81} Although an anisotropy of the resonance line
cannot be excluded due to the two-dimensional character of the
system, no such indications can be deduced from the powder spectra
in GdI$_2$. In general, powder spectra exhibit a characteristic
pattern due to the random distribution of the orientation of the
grains, if resonance field and linewidth are
anisotropic.\cite{Abragam70} The fact that the spectra are well
described by a single Dysonian line indicates that an anisotropy,
if existing, is smaller than the observed linewidth. As in the
present compound the linewidth $\Delta H$ is of the same order of
magnitude as the resonance field $H_{\rm res}$, both circular
components of the exciting linearly polarized microwave field have
to be taken into account. Therefore the resonance at the reversed
magnetic field $-H_{\rm res}$ has to be included into the fit
formula for the ESR signal given by
\begin{eqnarray}
&&\frac{dP}{dH} \propto \frac{d}{dH} \label{dyson}\\
&&\times\left\{\frac{\Delta H + \alpha (H-H_{\rm res})}{(H-H_{\rm
res})^2 + \Delta H^2} + \frac{\Delta H + \alpha (H+H_{\rm
res})}{(H+H_{\rm res})^2 + \Delta H^2}\right\} \nonumber
\end{eqnarray}
This is an asymmetric Lorentzian line, which includes both
absorption and dispersion, with $\alpha$ denoting the
dispersion-to-absorption ratio. Such asymmetric line shapes are
usually observed in semiconductors\cite{Ivanshin00} and
metals,\cite{Barnes81} where the skin effect drives electric and
magnetic microwave components out of phase and therefore leads to
an admixture of dispersion into the absorption spectra. For
samples small compared to the skin depth one expects a symmetric
absorption spectrum ($\alpha = 0$), whereas for samples large
compared to the skin depth absorption and dispersion are of equal
strength yielding an asymmetric resonance line ($\alpha = 1$). An
additional contribution to the asymmetry of the resonance line can
arise from the fact that $\Delta H$ is of the same order of
magnitude as $H_{\rm res}$, because then not only the overlap with
the resonance at $-H_{\rm res}$ but also the mutual coupling via
the nondiagonal elements of the dynamic susceptibility influences
the lineshape.\cite{Benner83} As can be seen in Fig.~\ref{Spectra}
the asymmetry of the resonance remains almost constant for higher
temperatures, while on approaching magnetic order from above the
resonance strongly shifts to lower fields and concomitantly the
asymmetry increases significantly due to the increasing overlap
with the corresponding resonance at $-H_{\rm res}$. The
temperature dependence of the parameter $\alpha$ is shown in the
inset of Fig.~\ref{XYZvsT}(b).

Figure \ref{XYZvsT} shows the temperature dependences of the main
spectral ESR parameters of GdI$_2$, namely (a) the intensity
$I_{\rm ESR}$, (b) the resonance field $H_{\rm res}$, and (c) the
half-width-half-maximum linewidth $\Delta H$, which will be
discussed in the following.

\subsection{ESR intensity and resonance field}

In principle, \iesr measures the spin susceptibility of the Gd
spins and is approximated by \iesr$\propto A \Delta H^2$, where
$A$ denotes the amplitude of the field-derivative of the
absorption spectrum. In metals, however, one has to account for
the finite skin depth $\delta$ of the microwave into the sample,
which depends on the resistivity $\rho$ and the ESR frequency
$\omega$ as $\delta=(\rho/\mu_0 \omega)^{0.5}$. As for GdI$_2$
only the resistance $R$ but not the specific resistivity $\rho$ is
available at present, the absolute skin depth cannot be
calculated. Note that the resistance in the paramagnetic regime
exhibits a broad maximum,\cite{Ahn00} and, therefore, the changes
in the skin depth are negligible, so that one probes always the
same fraction of the sample. As we are only interested in relative
changes of the spin susceptibility, the use of the above
approximation for \iesr is justified. \iesr decreases monotonously
with increasing temperature. Following the analysis of spin
fluctuations of localized  4$f^7$ electrons by Eremin \etal
\cite{Eremin01} for $T>T_{\rm C}$, we use the
random-phase-approxomation-like approach to the dynamic spin
susceptibility (depending on the wavevector $\mathbf{q}$ and the
frequency $\omega$) given by
\begin{equation}\label{spinsusc}
\chi_{\rm spin}(\mathbf{q},\omega)=\frac{\chi_{2\mathrm
D}(\mathbf{q})}{1-J_{\perp}(\mathbf{q})\chi_{2\mathrm
D}(\mathbf{q})}\frac{i\gamma(\mathbf{q},\omega)}{\omega+i\gamma(\mathbf{q},\omega)},
\end{equation}
where $\gamma(\mathbf{q},\omega)$ constitutes the damping function
of the spin fluctuations, and $J_{\perp}(\mathbf{q})$ the
ferromagnetic inter-layer exchange constant. Evaluation of the
imaginary part of Eq.(\ref{spinsusc}) at $\mathbf{q}=0$ and using
the temperature dependence of the 2D spin susceptibility
$\chi_{2\mathrm D}$ given by
\begin{equation}\label{2dsusc}
\chi_{\rm 2\mathrm
D}(T)=\frac{1}{6\sqrt{3}SJ_{\parallel}}\exp{\left[\frac{16\sqrt{3}S^2J_{\parallel}}{T}\right]},
\end{equation}
results in
\begin{equation}\label{fitformula}
\chi_{\rm spin}(T)\propto
\left(\exp{\left[-\frac{16\sqrt{3}S^2J_{\parallel}}{T}\right]}-\frac{J_{\perp}}{6\sqrt{3}SJ_{\parallel}}\right)^{-1},
\end{equation}
with the Gd spin $S=7/2$ and the ferromagnetic exchange constant
$J_{\parallel}$ within the Gd layers. Note that the determination
of absolute values of the susceptibility using
Eq.~(\ref{spinsusc}) cannot easily be undertaken, as the 'form of
the function $\gamma(\mathbf{q},\omega)$ for finite $\omega$ is
unknown'.\cite{Eremin01}

\begin{figure}[b]
\centering
\includegraphics*[width=70mm,clip]{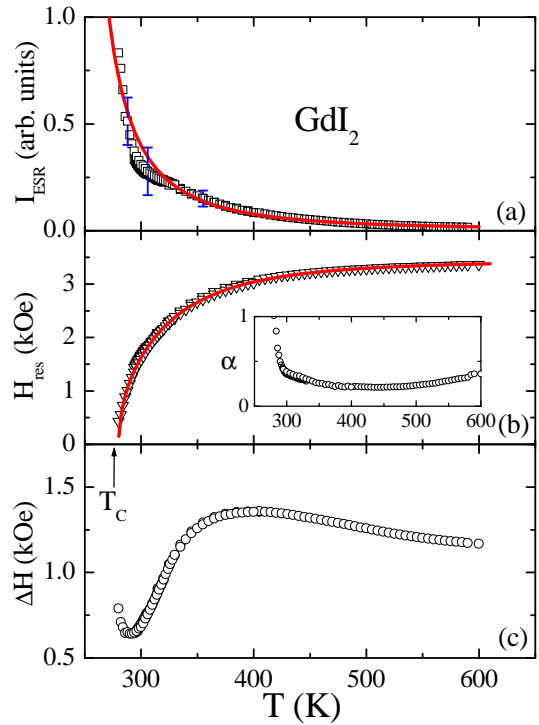}
\vspace{2mm} \caption[]{\label{XYZvsT}Temperature dependence of
(a) the ESR intensity $I_{\rm ESR}$, (b) the resonance field
$H_{\rm res}$ including the dispersion-to-absorption ratio
$\alpha$ in the inset, and (c) the ESR linewidth $\Delta H$ in
GdI$_2$. Solid lines in (a) and (b) represent fits obtained by
using Eqs.~(\ref{fitformula}) and (\ref{fieldfit}) as described in
the text, respectively.}
\end{figure}

The best fit obtained using Eq.~(\ref{fitformula}) is displayed as
a solid line in Fig.~\ref{XYZvsT}(a). Within experimental
uncertainties, the data can be well described by this approach
with fit parameters $J_{\parallel}=4.9(1)$ K and
$J_{\perp}/J_{\parallel}=0.035(3)$. These values are in good
agreement with the estimates $J_{\parallel}\approx 6$ K and
$J_{\perp}/J_{\parallel}\approx 0.03$ obtained from the analysis
of the Curie temperatures by Eremin and coworkers.\cite{Eremin01}
Deviations from the fitting curve are largest in the temperature
range $300<T<350$ K, where the linewidth is reduced by a factor of
three. The approximation \iesr$\propto A \Delta H^2$  strictly
holds only for $\Delta H\ll H_{\rm res}$ but can also be used for
larger $\Delta H$, if $\Delta H$ changes only slightly as in our
case above 350 K. Therefore, the largest uncertainty in \iesr
certainly occurs where the changes in linewidth are largest.

The resonance field \hres or the effective $g$-value $g_{\rm
eff}$=$h\nu/(\mu_{\rm B}H_{\rm res})$ provides information about
the local static magnetic field at the Gd site. The asymptotic
high-temperature value of the resonance field $H_{\rm
res}(\infty)=3.4$ kOe corresponds to an effective $g_{\rm
eff}$-value of $g=1.99$ similar to the value observed for
Gd$^{3+}$ ions in insulators,\cite{Buckmaster72} as expected for
vanishing orbital moment ($L=0$) of the half-filled 4$f$ shell. At
about 400 K the resonance field starts to decrease dramatically
towards the ferromagnetic ordering at $T_{\rm C}=276$
K.\cite{Ahn00} Considering the strong influence of the
ferromagnetic fluctuations on the resonance shift on approaching
$T_{\rm C}$ from above and the quite large values of the
susceptibility $\chi>0.1$ emu/mol for $T<400$ K,\cite{Eremin01}
demagnetization effects cannot be neglected. Due to the relation
between intrinsic and observed susceptibility
\begin{equation}
\frac{1}{\chi_{\mathrm{obs}}}=\frac{1}{\chi_{\mathrm{int}}}+N,
\label{referee2}
\end{equation}
where $N$ denotes the demagnetization factor of the order of
4$\pi$ along the direction of the applied magnetic field, one
expects important corrections. Therefore we use the famous formula
derived by Kittel\cite{Kittel48} for ferromagnetic resonance in an
ellipsoidal sample given by
\begin{equation}
\frac{\omega_0}{\gamma}= \sqrt{\left[H_{e0}+\left(N_x -N_z
\right)M_0 \right] \left[H_{e0}+\left(N_y -N_z \right)M_0
\right]}, \label{kittel}
\end{equation}
where $\omega_0$ denotes the resonance frequency, $\gamma$ the
gyromagnetic ratio, $H_{e0}$ the external magnetic field applied
parallel to the $z$-direction, $N_i$ the demagnetization factors,
and $M_0$ the magnetization of the sample. Approximating the shape
of the sample as a cylindrical disk with finite thickness ($N_y
=N_z \ll N_x \mbox{,}\quad N_x -N_z =:N_{\rm res}$) and with the
external magnetic field $H_{e0}$ applied within the disk's plane
one derives
\begin{equation}
H_{e0}= H_{0}/\sqrt{1+N_{\rm res}\chi}
\end{equation}
by using $M_0= \chi H_{e0}$ and $H_0=\omega_0/\gamma$. Applying
the proportionality for $\chi_{\rm spin}(T)$ given by
Eq.~(\ref{fitformula}) the following fit formula for the resonance
field $H_{\rm res}=H_{e0}$ emerges:
\begin{equation} \label{fieldfit}
H_{\rm res}(T)=
\frac{H_{0}}{\sqrt{1+A\left(\exp{\left[-\frac{16\sqrt{3}S^2J_{\parallel}}{T}\right]}-\frac{J_{\perp}}{6\sqrt{3}SJ_{\parallel}}\right)^{-1}}}
\end{equation}
The dimensionless parameter $A$ includes the demagnetization
factor and the unknown prefactors of the susceptibility (see
above). The resulting fit curve (solid line in
Fig.~\ref{XYZvsT}(b)) yields parameters $A=2.2(1)\times 10^{-3}$,
$H_0=3468(3)$ Oe, $J_{\parallel}=5.6(3)$ K, and
$J_{\perp}/J_{\parallel}=0.039(2)$. This good agreement with the
ESR intensity and the reported values by Eremin and
coworkers\cite{Eremin01} corroborates the consistency of our
approach.

\subsection{Linewidth}

\begin{figure}[h]
\centering
\includegraphics*[width=85mm,clip]{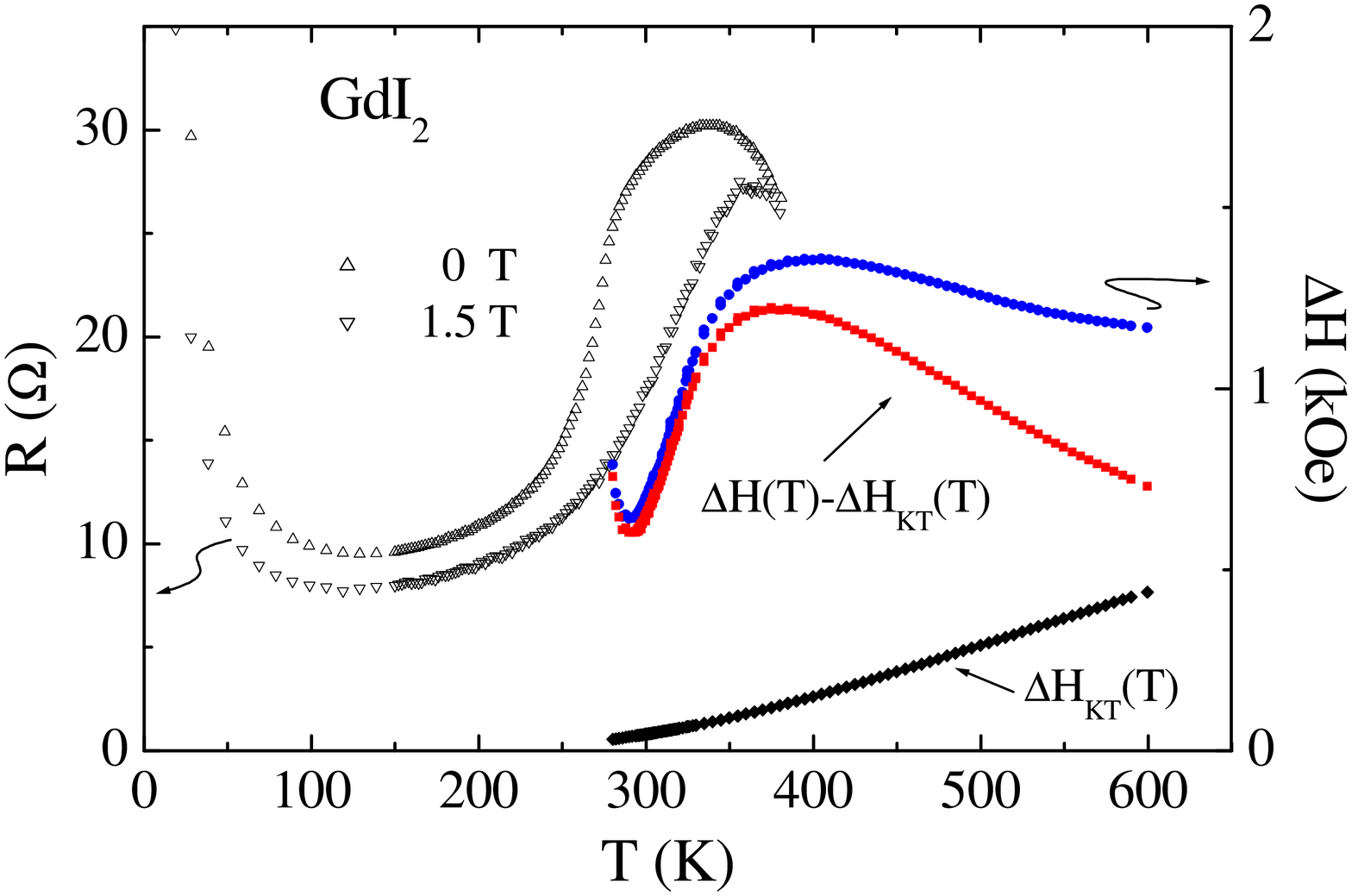}
\vspace{2mm} \caption[]{\label{dHvsT}Temperature dependence of the
ESR linewidth  together with the temperature dependence of the
resistance for GdI$_2$ in zero-magnetic field and in an applied
field of 1.5 T taken from Ref.~\onlinecite{Felser99}. The
contribution $\Delta H_{\rm KT}$ has been estimated as described
in the text and subtracted from the linewidth data to highlight
the contribution of the $d$-$f$ scattering.}
\end{figure}

On decreasing temperatures the resonance broadens and the
linewidth exhibits a broad maximum at about 390 K (see
Fig.~\ref{XYZvsT}(c)). Towards lower temperatures the linewidth
decreases and shows a minimum before it abruptly increases again
due to inhomogeneous broadening above the ferromagnetic
transition. Towards the highest temperatures the linewidth seems
to reach a constant value at about 1.1 kOe. Already at first sight
one recognizes that $\Delta H(T)$ correlates with the temperature
dependence of the electrical resistance, both of which are shown
in Fig.~\ref{dHvsT}. It has been proposed that the maximum in
$R(T)$ results from strong scattering of the $5d$-derived
conduction electrons at the localized $4f^7$-electrons due to a
peculiar topology of the Fermi surface in this
compound.\cite{Felser99} The conduction electrons in \GdI occupy a
narrow $5d_{z^2}$ band and the carriers at the Fermi level are
close to a $\Gamma$ point of the first Brillouin zone. Hence,
critical ferromagnetic fluctuations influence the magnetic part of
the resistivity very strongly.

The ESR linewidth $\Delta H(T)$ generally corresponds to the
transverse spin relaxation rate $1/T_2$ due to local fluctuating
fields, which in insulators originate from anisotropic
interactions like dipole-dipole interaction between the localized
spins, hyperfine fields from the nuclei, and the crystal electric
field of the surrounding ligands. In good metals, which contain
both localized magnetic moments and delocalized conduction
electrons, the interaction between them provides an additional
important relaxation channel that usually dominates the broadening
mechanism. Regarding the resistance values, GdI$_2$ cannot be
looked upon as a good metal and, hence, we will discuss both
line-broadening contributions in the following:

In exchange coupled systems the isotropic Heisenberg exchange $J$
(here $ J_{\parallel}\approx 6$ K) leads to an exchange narrowing
of the ESR spectrum into a single line, which is usually described
by the Kubo-Tomita approach,\cite{Kubo}
\begin{equation}
\Delta H_{\rm KT}\simeq\frac{M_2}{H_{\rm ex}} \label{narrowed}
\end{equation}
with the second moment $M_2$ of the resonance line and exchange
field $H_{\rm ex}\propto J$. The temperature dependence of the
linewidth in this case is given in the high-temperature
approximation by
\begin{equation}
\Delta H_{\rm KT}(T) = \frac{\chi_0(T)}{\chi(T)}\Delta H_{\infty}
\label{Huber}
\end{equation}
with the free Curie susceptibility $\chi_0=C/T$, where $C=N
g^2\mu^2_{\rm B} S(S+1)/3k_{\rm B}$ denotes the Curie constant of
the Gd ions with $S=7/2$, and the measured $dc$-susceptibility
$\chi(T)$.\cite{Huber71} As $\chi_0(T)/\chi(T)\rightarrow 1$ for
$T\rightarrow\infty$, the temperature independent parameter
$\Delta H_{\infty}$ can be identified with the high-temperature
limit of the ESR linewidth. To estimate this contribution to the
ESR linewidth, we follow the arguments by Sperlich \etal for the
case of GdB$_6$.\cite{Sperlich72} The authors conclude that the
main contribution to the resonance linewidth is given by the
dipole-dipole interaction of the rare-earth spins contributing
about 0.3 kOe to the linewidth for GdB$_6$, which has an exchange
constant of about 1 K. With $J\approx 6$ K in GdI$_2$ this value
is about 50 Oe and too small to fully account for the observed
high-temperature value of the linewidth. However, it cannot be
excluded that the strong two-dimensionality of the system
effectively reduces the exchange-narrowing effect resulting in a
larger contribution of the dipole-dipole
interaction.\cite{Richards74}

Due to very short correlation times $\tau\ll\omega_0$ in metals,
longitudinal $T_1$ and transversal spin relaxation time $T_2$ are
equal.\cite{Slichter} Hence, in this case the ESR linewidth
directly measures the spin-lattice relaxation rate 1/$T_1$. In the
case of a half-filled $4f$-shell with spin $J=S=7/2$ the orbital
momentum of Gd$^{3+}$ vanishes and a direct relaxation to the
lattice is not possible. Nevertheless, the energy is transferred
to the lattice by scattering of the conduction electrons at the
localized moment of the 4$f$ shell. This is the well-known
Korringa relaxation which gives rise to a linear increase of the
Gd linewidth with increasing temperature.\cite{Korringa} This
behaviour is typically observed for Gd ions diluted in usual
metals and also in many concentrated Gd-based transition-metal
compounds above magnetic order.\cite{Taylor75,Elschner97} The
slope of the linear increase is determined by the electronic
density of states at the Fermi level $\propto N^2(E_{\rm F})$.
Deviations from this canonical behaviour have already been found
from strongly correlated electron systems like heavy-fermion
compounds, where the density of states is strongly modified near
$E_{\rm F}$.\cite{KvN03} In these compounds, both the resistivity
and Gd-ESR linewidth show a typical non-linear temperature
dependence with a maximum near the characteristic temperature
$T^*\ll E_{\rm F}/k_{\rm B}$ for the screening of the Kondo ions
by the conduction electrons.

In GdI$_2$ we do not have any Kondo effect with its
antiferromagnetic screening of the localized moments,\cite{Kondo}
but instead there are strong ferromagnetic fluctuations due to the
exchange interaction between the Gd 5$d$ band and the localized
4$f$ shell. Like in the case of the Kondo effect the electrical
resistance is dominated by these fluctuations and vice versa the
spin relaxation of the Gd 4$f$ shell is strongly affected.
Therefore the maxima of resistance and linewidth which coincide
approximately indicate the characteristic temperature of the
ferromagnetic fluctuations in the range $T_{\rm C}<T^*<\Theta_{\rm
CW}=412$ K. The Curie-Weiss temperature $\Theta_{\rm CW}$ was
obtained from the high-temperature regime of the susceptibility
reported in Ref.~\onlinecite{Eremin01}.

To accentuate the contribution due to the ferromagnetic spin
fluctuations it seems useful to subtract the contribution of the
high-temperature relaxation. Both possible contributions to the
high-temperature linewidth, dipole-dipole interaction following
Eq.~(\ref{Huber}) or a linear Korringa increase, if metallic
properties dominate, would yield a monotonic increase with
increasing temperature. Subtracting such a contribution results,
hence, in a more pronounced maximum in agreement with the maximum
of the resistance. To illustrate this we assumed a
high-temperature value $\Delta H_{\infty}=1.1$ kOe corresponding
to the linewidth at the highest experimental temperatures and
simulated the Kubo-Tomita  contribution Eq.~(\ref{Huber}) using
the susceptibility data from Ref.~\onlinecite{Eremin01}.

Compared to $R(T)$, however, the linewidth maximum is shifted to
higher temperatures. This can be attributed to the finite external
magnetic field in the ESR experiment, which is swept over a range
from 0 to 1.5 T. Field-dependent resistance measurements evidence
a significant shift of the maximum to higher temperatures with
increasing strength of the magnetic field.\cite{Ahn00} From
theoretical estimates, it follows that the maximum in $R(T)$
shifts by approximately 10 \% in external magnetic fields of 1
T,\cite{Eremin01} hence accounting for the observed differences
between ESR linewidth and resistance.

Interestingly, a correlation between the temperature dependence of
$\Delta H(T)$ and $R(T)$ has also been observed in the
intermetallic compound Gd$_2$PdSi$_3$.\cite{Deisenhofer03} Similar
to \GdI this compound is  characterized by a large
magnetoresistance effect and, though showing antiferromagnetic
ordering at $T_{\mathrm N}=21$~K, exhibits a positive Curie-Weiss
temperature $\Theta_{\rm CW}=25$ K evidencing the existence of
strong ferromagnetic correlations.\cite{Mallik98a} On approaching
$T_{\mathrm N}$ both resistivity and ESR linewidth attain a
minimum at $T \approx 2T_{\mathrm N}$, followed by a similar
increase of both quantities towards the N\'{e}el temperature. It
seems likely that this feature could be due to similar scattering
processes as in GdI$_2$, especially, because the occurrence of
anisotropic magnetoresistance effects and anomalies in the Hall
resistivity strongly suggest the existence of an anisotropic Fermi
surface and its reconstruction across the metamagnetic anomaly
observed in Gd$_2$PdSi$_3$.\cite{Saha99} To establish such a
scenario, however, a theoretical analysis of the ferromagnetic
correlations in Gd$_2$PdSi$_3$ is highly desirable.

\section{Conclusions}
In summary, we find that in \GdI the temperature dependence of the
ESR linewidth correlates with the resistance and conclude that the
linewidth is dominated by scattering processes of conduction
electrons at the localized $f$ electrons. Both the resonance field
and the ESR intensity can be well described by a phenomenological
approach taking into account ferromagnetic spin fluctuations of
the $f$ electrons. For the resonance field and the ESR intensity
the analysis yields in- and out-of-plane exchange constants
$J_{\parallel}=5.6;4.9$ K and
$J_{\perp}/J_{\parallel}=0.039;0.035$, respectively, which are in
excellent agreement with previous estimates.

\begin{acknowledgments}
We thank I.~Eremin for fruitful discussions. This research was
supported by the Bundesministerium f{\"u}r Bildung und Forschung BMBF
via the contract number VDI/EKM 13N6917 and partly by the Deutsche
Forschungsgemeinschaft DFG via SFB 484 (Augsburg).
\end{acknowledgments}


\end{document}